\documentstyle[prd,aps,preprint]{revtex}

\tighten

\begin{document}
\draft

\pagestyle{empty}

\preprint{
\noindent
\hfill
\begin{minipage}[t]{3in}
\begin{flushright}
LBNL-41255\\
UCB-PTH-98/02\\
January 1998\\
\end{flushright}
\end{minipage}}

\title{Long-distance final-state interactions and J/$\psi$ decay}

\author{
Mahiko Suzuki}
\address{
Department of Physics and Lawrence Berkeley National Laboratory\\
University of California, Berkeley, California 94720}

\maketitle

\begin{abstract}

To understand the short versus long-distance
final-state interactions, we have performed
a detailed amplitude analysis for the two-body decay 
$J/\psi\rightarrow 1^-0^-$. The current data favor 
a large relative phase nearly $90^{\circ}$ between the three-gluon
and the one-photon decay amplitudes. The source of the phase is apparently
the long-distance final-state interaction.  Nothing anomalous is found in 
the magnitudes of the three-gluon and one-photon decay amplitudes.
We discuss the implications of this large phase 
in the weak decay of heavy particles.
\end{abstract}


\pagestyle{plain}
\narrowtext

\section{Introduction}
     
   Though the final-state interaction phases are important to 
observability of CP violating decays, it is very difficult to compute them
or to extract them from data. Only the short-distance contribution has been 
computed in the quark-gluon picture\cite{Soni,Kramer}. Any attempt to estimate 
the long-distance contribution in the hadron picture has so far been 
limited to elastic or quasi-elastic rescattering, which is 
presumably only a small portion of the long-distance effect, 
particularly in the heavy particle decays like B-decay.    
Some argue that the long-distance contribution is negligible
at least in some decay modes\cite{BJ}, while others identify 
a specific long-distance contribution and show that it is 
actually much larger than the short-distance effect\cite{Donoghue}. 
Some entertain the idea that many long-distance contributions might 
average out to a small effect after being summed up.    

   In this paper we try to test whether the short-distance final-state 
interaction dominates over the long-distance one
in the $J/\psi$ decay or not.  Though the $J/\psi$ decay proceeds
with strong and electromagnetic interactions, its narrow width allows 
us to treat the decay just like weak decays, namely, a short-distance decay 
followed by long-distance rescattering.  In the quark-gluon picture 
$J/\psi$ decays either directly into three gluons or into a quark
and an antiquark through one photon. Both processes acquire 
a short-distance QCD rescattering phase of $O(\alpha_s/\pi)$.  
If this is the dominant source of the final-state interaction phase, 
the relative phase ought to be very small for the amplitudes of all
decay modes. On the other hand, if long-distance processes are important to
generating the phase, the decay amplitudes can have large relative 
phases to each other.  Since there are sufficient data on the decay 
$J/\psi\rightarrow 1^-0^-$, we are able to perform an amplitude analysis 
and extract the relative phase between the three-gluon decay and 
the one-photon decay amplitudes.     

      In Section II we attempt a detailed numerical analysis of the decay 
amplitudes in the framework of the broken flavor SU(3) symmetry. We include 
all first-order symmetry breakings and some of second order effects.  
The result of our analysis shows that the three-gluon and one-photon amplitudes
have a large relative phase of rescattering, nearly $90^{\circ}$
off phase to each other.
It indicates that the major source of the final-state interaction phases 
is in the long-distance hadronic rescattering.

      The decay $J/\psi\rightarrow 1^-0^-$ has been a subject of
discussion in connection with the abnormally small yield of the decay 
$\psi'\rightarrow 1^-0^-$. A few exotic models were proposed to resolve
this puzzle\cite{Hou,BLT,BK}. They suggested that the dominant process of 
$J/\psi\rightarrow 1^-0^-$ is not a perturbative three-gluon decay. However 
the result of our analysis shows that the magnitude of the $I=0$ 
decay amplitudes is consistent with the three-gluon decay. We can also show 
a serious shortcoming of those models.  We shall discuss 
on this point in Section III.  Finally in Section IV, we 
discuss on the implications of the present analysis in the B-decay.    

\section{Amplitude analysis in broken flavor SU(3) symmetry}

    Before entering our amplitude analysis, we read off
one relevant information from the current data.  
The lepton-pair decay branching fraction $B_l$ for 
$J/\psi\rightarrow\gamma\rightarrow l^+l^-(=e^+e^- + \mu^+\mu^-)$
has been measured with a high accuracy. In contrast, 
the inclusive one-photon annihilation into hadrons for $J/\psi
\rightarrow\gamma\rightarrow q\bar{q}(= u\bar{u}+d\bar{d}+s\bar{s})$
was obtained only indirectly from the nonresonant background cross section
$\sigma(e^+e^-\rightarrow q\bar{q})$ interpolated to the $J/\psi$ mass.  This
two decades old value\cite{MarkI} quoted in the Review of Particle Physics 
(RPP)\cite{RPP} has a large uncertainty. Actually we can obtain a more
accurate value for $B_{\gamma}\equiv B(J/\psi\rightarrow\gamma\rightarrow
q\bar{q})$ by making the perturbative correction of 
$(1+\overline{\alpha}_s/\pi)$ to the leptonic branching fraction $B_l$. 
With this help of theory, we can compute the inclusive three-gluon 
decay branching fraction $B_{ggg}$ with $B_{ggg}=1-B_l-B_{\gamma}$.
We thus obtain for the inclusive three-gluon and one-photon 
annihilations into hadrons
\begin{equation}
   \frac{B_{ggg}}{B_{\gamma}} = 5.8 \pm 0.3 \; (4.2 \pm 0.6),
                                               \label{inclusivebranching}
\end{equation}
where the number in the bracket is the {\it experimental} value quoted in 
RPP. Except for the number in Eq.(\ref{inclusivebranching}), we take the 
numbers tabulated in RPP as the {\it current} data throughout this paper.    

If branching fractions of the $J/\psi\rightarrow 1^-0^-$ modes scale with
those of the inclusive hadron decay processes, we expect that the amplitude
analysis should give us the ratio of the three-gluon to one-photon decay 
amplitudes into the exclusive channels $1^-0^-$  somewhere between 
2 and 3 ($\approx\sqrt{5.8}$).  Later we shall often compare the results of 
our amplitude analysis with this number.

\subsection{Fit to $J/\psi\rightarrow V_9 P_8$}

   We first study the two-body decay modes of $J/\psi\rightarrow 1^-0^-$ in
which the pseudoscalar meson belongs to an SU(3) octet. The singlet $\eta'$ 
will be included later with the $\eta-\eta'$ mixing.  
For the vector mesons, we study the singlet and 
the octet together as a nonet. We parametrize the decay amplitudes as
follows:

(1)  The $1^-$ mesons form the ideally mixed nonet\cite{nonet}, namely 
$\phi = -\bar{s}s$. Therefore the SU(3) symmetric coupling is given by  
\begin{equation}
    L_{int}= a\;{\rm tr}(V_9 P_8),
\end{equation}
where $V_9$ and $P_{8}$ are represented in the $3\times 3$ matrices.

(2)  The strong SU(3) breaking of $\lambda_8$ is included in the three-gluon
decay:
\begin{equation}
   L_{int}= \epsilon\;{\rm tr}(\{V_9,P_8\}_{+} T_3^3),
\end{equation}
where we use $T_3^3$ instead of $\lambda_8$ to simplify the numerical
coefficients of parametrization.
The symmetrization of $V_9$ and $P_8$ is required by charge conjugation 
invariance.
 
(3)  The one-photon annihilation amplitudes transform like $\lambda_{E}
=(\lambda_3 +\lambda_8/\sqrt{3})/2$. Therefore, they are parametrized as:
\begin{equation}
   L_{int}= a_{\gamma}{\rm tr}(\{V_9, P_8\}_{+}\lambda_{E}).
\end{equation}

(4)  The phases of the amplitudes $a, \epsilon$, and $a_{\gamma}$ are group
theoretically independent.  Therefore, we introduce two relative phases,
$\delta_{\gamma}$ and $\delta_{\epsilon}$, defined as: 
\begin{equation}
               {\rm arg}(a_{\gamma}a^*) = \delta_{\gamma}\;\;\;
               {\rm arg}(\epsilon a^*) = \delta_{\epsilon}.
\end{equation}

(5) The $\rho-\omega$\cite{rho-omega} mixing can be potentially important to 
the processes of $\Delta I = 1$.  For instance, the process
$J/\psi\rightarrow\omega\eta\rightarrow\rho\eta$ 
interferes with the direct process $J/\psi\rightarrow\gamma\rightarrow\rho
\eta$ and is counted as part of the branching to $J/\psi\rightarrow\rho\eta$. 
The decay $J/\psi\rightarrow\omega\eta$ can proceed through 
three gluons while the direct process is the one-photon process. Therefore,
the $\rho-\omega$ mixing may not be negligible in this mode. Since the
$\omega$ width is much narrower than the $\rho$ width, the major
contribution of the $\rho-\omega$ transition to $J/\psi\rightarrow\rho\eta$
occurs at the $\omega$ resonance peak of $\pi^+\pi^-$. 
In comparison, in the process like $J/\psi\rightarrow\omega\pi^0$, 
the effect of the $\rho-\omega$ transition is less significant because 
the large $\rho$ width suppresses it kinematically. In the processes 
of $\Delta I = 0$, the $\rho-\omega$ transition is negligible.
  
We have included the $\rho-\omega$ mixing with the
effective transition coupling,

\begin{equation}
   L_{int} = m_{\omega}^2 f_{\rho\omega}\rho_{\mu}\omega^{\mu},
\end{equation}  
with $f_{\rho\omega}= - 6.8\times 10^{-3}$, as extracted from the
process $e^+e^-\rightarrow\pi^+\pi^-$ around the $\omega$ mass.

(6) The p-wave phase space correction is made with $p^3$ for all decay
branching fractions. If the flavor symmetry applies best to the dimensionless
decay couplings, the phase space factor $p^3$ should be divided by some
quantity having the dimension of squared mass that may be subject to  
symmetry breaking.  However, this uncertainty has been incorporated 
through the $\lambda_8$ breaking of (2) above, at least, to the first order.  

   In Table I, we have tabulated the parametrization of seven 
$V_9P_8$ decay amplitudes with the SU(3) amplitudes,
\begin{equation}
             a,\; \epsilon, \; a_8.   \label{3parameters}
\end{equation}
If the $1^-0^-$ decay branchings scale more or less with the inclusive ones
({\it cf} Eq.(\ref{inclusivebranching})), we expect $|a_{\gamma}/a|\approx 
0.67$ in the normalization used in Table I.  For the strong symmetry breaking,
$|\epsilon/a|\leq 0.3$ is a reasonable range.
In addition, we have two relative phases as free parameters,
$ \delta_{\epsilon}$ and $ \delta_{\gamma}$.

The best fit with these five parameters is obtained for
\begin{equation}
         a=1,\; \epsilon =-0.22 ,\; a_{\gamma}= 0.34,\; 
         \delta_{\epsilon}=-22.5^{\circ} , \; \delta_{\gamma} =80.3^{\circ} ,
                 \label{7fit}
\end{equation}
where magnitude of the amplitude $a$ has been normalized to unity. The fitted 
values to the data are tabulated in the third column of Table next to 
the observed values. The $\chi^2$ is 4.8 for this fit.  
Though the ratio of $a_{\gamma}/a$ in Eq.(\ref{7fit}) is a half of 
the scaled value $\approx 0.67$, it can hardly be called an enhancement of
the three-gluon decay.  The magnitudes of
$a, a_{\gamma}$, and $\epsilon$ are in line with the expectation from the 
inclusive branching ratios: There is no sign of significant enhancement 
of the three-gluon processes relative to the one-photon processes.

   If we fit the data without the phases, the best $\chi^2$ is 57. When there
is no rescattering phase, the $\rho-\omega$ mixing is unimportant because the 
$\rho-\omega$ transition amplitudes are $90^{\circ}$ off phase to the main
amplitudes.  The fitted values are listed in the fourth column (No phase I) 
of Table II.
  In order to make a quantitative comparison of the fits with and without 
the phases, it is more appropriate to fit the data with the same 
number of free parameters, namely five real amplitudes.  We may add the
amplitudes of the $\lambda_8$ breaking to $a_{\gamma}$ as the 
second-order small quantities. 
   We may also include breakdown of the ideal nonet coupling ansatz. Actually, 
there is a subtlety between breakdown of the nonet and the $O(\lambda_8 e)$ 
correction. In the nonet scheme, which is realized in the nonrelativistic 
quark model, $V_1$ does not form an SU(3) coupling 
without accompanying $V_8$. For instance, the term like 
${\rm tr}V_9 \propto V_1$ is not allowed in an SU(3) coupling. 
When a $\lambda_8$ breaking is taken into account, however, 
one may include ${\rm tr}(\lambda_8 V_9)$ or ${\rm tr}(T_3^3 V_9)$ among others.
Normally it would not matter which of these is included since their difference
has the same SU(3) property as the term of the symmetry limit.  In the case of
the nonet, the difference $\sim{\rm tr}V_9$ is a term forbidden by the nonet
coupling ansatz. Therefore we must choose between them either from the 
observed $V_9$ parameters or from some theoretical reasoning. Since  
the $\lambda_8$ breaking is caused by the s-quark mass term and the
orgin of the nonet is in the nonrelativistic quark model, we feel that 
$T_3^3$ is more appropriate than $\lambda_8$ in the case of the nonet.
We include the strong SU(3) breaking to $a_{\gamma}$ along this line to 
see an outcome of the fit. 
     
After charge conjugation invariance 
is taken into account, there are three independent amplitudes 
of O($T_3^3 e$) that have different group structures 
from the amplitudes in Eq.(\ref{3parameters}). One of them ($\propto 
{\rm tr}(T_3^3 V_9)$) contributes only to the $\phi \pi^0$ and $\phi\eta$ 
modes. Meanwhile a severe upper bound has been set
on $B(J/\psi\rightarrow\phi\pi^0)$ by experiment. With this 
upper bound, we find that this SU(3) amplitude contributes no more than 
$6\%$ to $\phi\eta$, smaller than normally expected for strong SU(3) 
breakings and below the level of our concern.  We therefore drop this small
amplitude and retain only the remaining two amplitudes as 
the O($T_3^3 e$) corrections:
\begin{equation}
     L_{int}= \epsilon_{\gamma 1}[{\rm tr}(V_9\lambda_{E} P_8 T_3^3)
                                 +{\rm tr}(P_8\lambda_{E} V_9 T_3^3)]
           +\epsilon_{\gamma 2}{\rm tr}(P_8 T_3^3){\rm tr}(V_9\lambda_{E}).
\end{equation}
Now we have five real amplitudes.  The best fit attains $\chi^2=21$ 
in this case.   The fitted values are:
\begin{equation}
   a = 1, \; \epsilon =-0.14 , \; a_{\gamma}=0.30 , \;
   \epsilon_{\gamma 1}=-0.12   , \; \epsilon_{\gamma 2}= -0.11 .  
\end{equation}
The fitted branching fractions are listed in the fifth column 
(No phase II) of Table II.

A simple qualitative explanation can be given as to 
why the best fit needs the phases. Refer to the Table I and the 
observed branching fractions listed in Table II for the following discussion:

1. First of all, the significant difference between the $\rho\pi$ and 
$K^{+*}K^-$ branchings must be explained by the SU(3) breaking
$\epsilon$ amplitude. The $\epsilon$ amplitude of a right magnitude
($|\epsilon/a|=0.22$) produces this difference.

2. Next, we need a sizable $a_{\gamma}$ amplitude  
in order to account for the $\omega\pi^0$ mode.

3. The $a_{\gamma}$ amplitude contributes to splitting the branching fractions 
of $K^{+*}K^-$ and $K^{0*}\overline{K}^0$ too. However, if $a_{\gamma}$ and 
$a$ substantially interfered, this splitting would be much too large. To keep
the splitting between $K^{+*}K^-$ and $K^{0*}\overline{K}^0$  small,
$a_{\gamma}$ and $a$ must be largely off phase to each other.

The relative phase between $\epsilon$ and $a$ turns out to be small. 
One interpretation for the smallness of this relative phase is 
that the main source of $\epsilon$ is the kinematical SU(3) 
breaking due to mass splitting in the phase space and decay coupling.

\subsection{Including $\eta'$}

   Once $\eta'$ is included, the number of independent 
parameters suddenly increases since unlike the vector meson couplings, 
the singlet $0^-$ couplings are not related to the 
octet $0^-$ couplings.  Complication grows further when we include the 
$\eta-\eta'$ mixing.

    Here we present a relatively simple sample of analysis instead of
the most general one in order to show that need of the relative phase between
the three-gluon and the one-photon amplitudes persists.  We add 
all SU(3) independent amplitudes involving $\eta'$ that correspond to
$a$, $\epsilon$, and $\a_{\gamma}$ of $P_8$:
\begin{equation}
   L_{int}= a'{\rm tr}(V_9 P_1) +\epsilon'{\rm tr}(V_9 P_1 T_3^3) 
            + a'_{\gamma}{\rm tr}(V_9 P_1\lambda_{E}).
\end{equation}
In addition to the $\rho-\omega$ mixing, we include the $\eta-\eta'$ mixing,
\begin{equation}
    \eta=\eta_8\cos\theta_p -\eta_1\sin\theta_p,\;\;
   \eta'=\eta_8\sin\theta_p +\eta_1\cos\theta_p,
\end{equation}
where $\theta_p$ is $-10^{\circ}\sim -20^{\circ}$\cite{etamixing}.
For the relative phase, we put a common phase $\delta$ between all one-photon
amplitudes and all three-gluon amplitudes ({\it e.g.,} $\delta_{\epsilon}=0$).

The best fit to the data is obtained with the following values of the
parameters:
\begin{eqnarray}
   a & = &1, \; \epsilon =-0.18, \; a_{\gamma} = 0.36, \\ \nonumber
   a'&=& 0.44,\; \epsilon' =0.051,\; a'_{\gamma}= -0.40, \\ \nonumber
   \delta &=& 75.2^{\circ},      \label{fit}
\end{eqnarray}
The value of $\chi^2$ is 8.2 for fitting  eleven data with seven 
parameters. The relative magnitudes of the parameters are {\it normal}, namely,
in line with the expectation from the inclusive branchings and strong SU(3) 
breakings. On the other hand, if we attempt to fit the data without the 
phase $\delta$, the $\chi^2$ jumps to 43.  In this case, sum of the one-photon 
branchings is close to that of the three-gluon branchings for $V_9P_1$. 
The tendency of deterioration of the fit without a phase
persists as we have seen in the case without $\eta'$.  
The fitted values of the branching fractions are tabulated with and 
without the phase in the last two columns of Table II.  The phase is 
unimportant in fitting to the decay modes involving $\eta'$. 

We conclude that as long as the currently listed data are taken at their 
face value, the three-gluon and one-photon amplitudes have a large relative
phase to each other. Apart from this unexpected result, our amplitude
analysis show that the magnitudes of all decay amplitudes are within the range
of what we expect.

\section{Is the decay $J/\psi\rightarrow 1^-0^-$ anomalous ?}

We have chosen the $J/\psi\rightarrow 1^-0^-$ decay modes for study of 
the final-state interaction phases since they are the most extensively 
measured decay modes.  No similar analysis can be made for other modes at
present.  

Meanwhile, there was one disturbing twist related to these decay modes. That is,
the $\psi'\rightarrow 1^-0^-$ decay modes are severely suppressed in comparison
with the corresponding $J/\psi$ modes\cite{anomaly}.
For $\rho\pi$, the upper bound on the branching fractions normalized 
to the $e^+e^-$ branching fraction obey the inequality,
\begin{equation}
     \frac{B(\psi'\rightarrow\rho\pi)}{B(\psi'\rightarrow e^+e^-)}
        < 1.2\times 10^{-2}\times
    \frac{B(J/\psi\rightarrow\rho\pi)}{B(J/\psi\rightarrow e^+e^-)}.
                                        \label{suppression}
\end{equation}
This vast difference between $J/\psi$ and $\psi'$ has stimulated many 
speculations on the pure QCD decay of $J/\psi$ and $\psi'$. 
The argument goes as follows:  
Normally $J/\psi(\psi')\rightarrow 1^-0^-$ would be highly suppressed 
by chirality mismatch of perturbative QCD.  However, this suppression 
is compensated by an enhancement in the $I=0$ channels of $J/\psi$.
The enhancement brings $\Gamma(J/\psi\rightarrow 1^-0^-)$ back to the value
predicted by the perturbative three-gluon annihilation\cite{BLT}.
The cause of enhancement may be either a vector gluonium state nearly 
degenerate with $J/\psi$\cite{Hou,BLT} or an hidden charm pair in the light 
$1^-$ mesons\cite{BK}.  $\Gamma(\psi'\rightarrow 1^-0^-)$ is small because
it suffers from chirality mismatch but receives no enhancement.

However, our amplitude analysis raises a doubt about such an
explanation. We have seen that both the $I=0$ and the $I=1$ 
amplitudes are as normal as we expect from the inclusive three-gluon and 
one-photon annihilations. If the observed magnitude of the $I=0$ 
amplitudes were actually the result of the compensation 
between a chirality suppression and a dynamical $I = 0$ 
enhancement, we would expect that the one-photon annihilation 
amplitude for $\omega\pi (I=1)$ should be suppressed by 
chirality without a compensating enhancement.  If their models
are correct, we can read off the chirality suppression factor 
from Eq.(\ref{suppression}). With the
chirality suppression, $B(J/\psi\rightarrow\omega\pi^0)$ would have to be
\begin{eqnarray}
          B(J/\psi\rightarrow\omega\pi^0) &\approx&
             ({\rm chirality}\; {\rm suppression})\times
             \frac{B(J/\psi\rightarrow\gamma\rightarrow q\bar{q})}
             {B(J/\psi\rightarrow ggg)}\cdot 
              B(J/\psi\rightarrow\rho\pi) \\ \nonumber
               &<& \frac{1}{200}\times B(J/\psi\rightarrow\rho\pi)
\end{eqnarray}  
in those models.  The data violate this inequality by 
two orders of magnitude.  Therefore, the origin of the
relative suppression of $\psi'\rightarrow 1^-0^-$ to $J/\psi\rightarrow
1^-0^-$ is not in $J/\psi$ but in $\psi'$. The large relative phase between 
$a$ and $a_{\gamma}$ cannot be attributed to a resonance in the $s$-channel 
of $I = 0$. Since the $I = 0$ amplitudes receive no net enhancement, 
a contribution of an $s$-channel resonance, if any, would be a tiny fraction 
of the whole amplitude\cite{Hou,Bai}.  Then the $I = 0$ amplitudes could
not have a large phase close to $90^{\circ}$.  It should be pointed out that
there exist other attempts to explain the relative suppression of
$\psi'\rightarrow 1^-0^-$ with different dynamical assumptions or intricate
dynamical coincidences\cite{Karl,Chaichian,Pinsky,Li}. The possibility of
a destructive interference in $\psi'$\cite{Li}, though it is fortuitous, 
cannot be ruled out in view of our finding of the large long-distance 
final-state interaction in $J/\psi$.
Whatever the cause of the $\psi'\rightarrow 1^-0^-$ suppression may be, the
group theoretical parametrization of the amplitudes remains the same for
$J/\psi\rightarrow 1^-0^-$. Though we are unable 
to choose the solution to this $\psi'\rightarrow\rho\pi$ puzzle among the 
existing models at present, we are confident that this puzzle does not 
interfere with our analysis in this paper. 

\section{Implication of the large relative phase} 

  In our analysis we have found the first evidence for a large final-state
interaction phase in a heavy particle decay which is quite different
in nature from the common subchannel resonant phases.
  What generates the large relative phase $\delta_{\gamma}$ between
$a_{\gamma}$ and $a$ ? It is obvious that it must arise from
long-distance strong interactions. The short-distance final-state interaction
phase difference can be evaluated in the quark-gluon picture of 
perturbative QCD. It is of
O($\alpha_s/\pi$) where $\alpha_s/\pi$ is 0.1 or less.  The large
phase difference close to $90^{\circ}$ found in our analysis
cannot be produced with the perturbative QCD interaction.  The source of  
$\delta_{\gamma}$ must be in the long-distance part of strong interactions,
namely, rescattering among hadrons in their inelastic energy region. 

When many channels are open for strong interaction rescattering, the phase 
of a decay amplitude into a physically observed state is determined by 
the phase shifts of eigenchannels of S-matrix and the coupling to them.
In the case of the $J/\psi$ decay, the decay amplitude $D(J/\psi\rightarrow h)$
into a final state $h$ ({\it e.g.,} $\rho^+\pi^-$) is written in the form
\begin{equation}
 D(J/\psi\rightarrow h) = \sum_{\alpha}X(J/\psi\rightarrow\alpha)
                     e^{i\delta_{\alpha}}O_{\alpha h}, \label{fsi}
\end{equation} 
where $\alpha$ refers to the eigenchannels of the partial-wave S-matrix of 
$J^{PC}=1^{--}$ at the energy of the $J/\psi$ mass with $\delta_{\alpha}$'s 
being their eigenphase shifts.  Time-reversal invariance requires that 
$X(J/\psi\rightarrow\alpha)^* =X(J/\psi\rightarrow\alpha)$ and that 
$O_{\alpha h}$ be an orthogonal matrix relating $h$ to $\alpha$ by $\langle h|
=\sum\langle\alpha|O_{\alpha h}$.  The partial-wave phase shifts 
contain much of long-distance physics no matter how high the energy is. 
One indication of substantial long-distance physics in the high-energy phase
shifts was pointed out by making a partial-wave projection of the diffractive
scattering amplitude\cite{Donoghue}. Long-distance physics enters the
eigenchannel matrix $O_{\alpha h}$ as well.
When we compare the three-gluon and one-photon 
decay amplitudes of $J/\psi$ with Eq.(\ref{fsi}), we see no simple 
relation between their phases in general: 
The eigenphase factors $e^{\delta_{\alpha}}$ are summed with the weights 
$X(J/\psi\rightarrow\alpha)$ different for $J/\psi\rightarrow ggg$ and
$J/\psi\rightarrow\gamma\rightarrow q\bar{q}$, leading to two phases
practically unrelated to each other.  In this picture the final-state
interaction phases of the $J/\psi$ decay are generally not determined by
short-distance physics alone even though pertubative QCD applies to the
inclusive $J/\psi$ decays.  The analysis of this paper indicates that 
long-distance physics can be far more important in the exclusive decays.

The conclusion of our amplitude analysis, if it is sustained, 
has a significant implication in a wide 
range of phenomena. For instance, when we evaluate the baryon asymmetry in the 
early Universe from CP-violating particle decays, 
we compute only the short-distance contribution of final-state interactions.  
Such a calculation makes sense only as an
order-of-magnitude estimate at best. In the case of the baryon 
asymmetry we may not ask for a high precision after all. 
However, in the B-meson decay where knowledge of much higher precision will 
be needed for final-state interaction phases, 
we shall have to know the long-distance final-state interaction phases
above the inelastic thresholds. It is nearly an impossible task to either
compute them theoretically or extract them from scattering data. 
If this is the case, the parameters of the fundamental interactions 
can be extracted only from those data which are free from complications 
due to the final-state interaction. It will not be an easy task
to look for meaningful physics in the rest of data.

  To conclude this paper, we should emphasize that numerical conclusion of
our analysis replies on the current data listed in RPP, not only their
central values but also the experimental uncertainties. We cannot rule out the
possibility that a future change in the data may upset our conclusion, {\it 
i.e.,} the need of a large rescattering phase.  For this reason, high
precision measurement of the $J/\psi$ decay branchings, particularly for
$\rho\pi$, $K^{*}\overline {K}$ and $\omega\pi^0$, will be very important
to our understanding of the final-state interactions in general.

\acknowledgements

This work was
supported in part by the Director, Office of Energy Research, Office of
High Energy and Nuclear Physics, Division of High Energy Physics of the
U.S.  Department of Energy under Contract DE--AC03--76SF00098 and in
part by the NSF under grant PHY--95--14797.

\begin{table}
\caption{The SU(3) parametrization of the $1^-0^-$ decay amplitudes.  
Listed are the coefficients of the amplitudes written on the top 
of each column. For instance, the coefficient of $a_{\gamma}$ for
$K^{*0}\overline{K}^0$ is $-2\times(1/3)$.
The $\eta-\eta'$ mixing and the $\rho-\omega$ mixing are 
introduced as explained in the text}

\begin{tabular}{|c|r|r|r|r|r|r|r|r|}
  Decay modes & $a$ & $ \epsilon$ & $\frac{1}{3} a_{\gamma}$ &
  $\frac{1}{3}\epsilon_{\gamma 1}$ & $\frac{1}{3}\epsilon_{\gamma 2}$&
  $a'$ & $\epsilon'$ & $\frac{1}{3}a'_{\gamma}$ \\ \hline
  $\rho^+\pi^-(=\rho^-\pi^+)$ &  1 &0& 1 &0&0&0&0&0\\ \hline
  $\rho^0\pi^0$               &  1 &0& 1 &0&0&0&0&0\\ \hline
  $K^{*+}K^-(=K^{*-}K^+)$     &  1 &1& 1 &2&0&0&0&0\\ \hline
  $K^{*0}\overline{K}^0(=\overline{K}^{*0}K^0)$
                              &  1 &1& -2 &-1&0&0&0&0\\ \hline 
  $\omega\pi^0$               &  0 &0& 3&0&0&0&0&0\\ \hline
  $\rho^0\eta_8$           &  0 & 0 &$\sqrt{3}$&0&$-\sqrt{3}$&0&0&0\\ \hline
  $\omega\eta_8$  &$\sqrt{1/3}$& 0 &$\sqrt{1/3}$&0&$-\sqrt{1/3}$&0&0&0\\ \hline
  $\phi\eta_8$    &$\sqrt{2/3}$&$\sqrt{8/3}$&$-\sqrt{8/3}$&
                   $-\sqrt{8/3}$&$-\sqrt{2/3}$&0&0&0\\ \hline
  $\rho^0\eta_1$   &0&0&0&0&0&$0$&0&$\sqrt{3/2}$\\ \hline
  $\omega\eta_1$   &0&0&0&0&0&$\sqrt{2/3}$&0&$\sqrt{1/6}$\\ \hline
  $\phi\eta_1$     &0&0&0&0&0&$-\sqrt{1/3}$&$-\sqrt{1/3}$&$\sqrt{1/3}$\\ \hline
  $\phi\pi^0$      &0&0&0&0&0&0&0&0
\end{tabular}
\label{table:1}
\end{table}
\begin{table}
\caption{The observed branching fractions and our fits with and without
$\eta'$ included. {\it No phase} means a fit assuming that all amplitudes
be real.  The bottom row lists $\chi^2$ for each fit. All numbers are 
in percent except for $\chi^2$.}

\begin{tabular}{|c|c|c|c|c|c|c|}
   Decay modes & Observed  & 
   \multicolumn{3}{c|}{$V_9P_8$} & 
   \multicolumn{2}{c|}{$V_9P_{8,1}$}\\ \cline{3-7} 
   &(in percent) &Best fit&No phase I&No phase II&Best fit&No phase\\ \hline
   $\rho^+\pi^-(=\rho^-\pi^+)$ &0.43$\pm$ 0.03 &0.44 & 0.44 & 0.47 
                             &0.43&0.42\\ \hline
   $\rho^0\pi^0$               &0.42$\pm$ 0.05 &0.44 & 0.44 & 0.47 
                             &0.43&0.42\\ \hline
   $K^{*+}K^-(=K^{*-}K^+)$     &0.25$\pm$ 0.02 &0.25 & 0.30 & 0.25 
                               &0.25&0.31\\ \hline
   $K^{*0}\overline{K}^0(=\overline{K}^{*0}K^0)$
                               &0.21$\pm$ 0.02 &0.21 & 0.14 & 0.16 
                               &0.21&0.16\\ \hline
   $\omega\pi^0$               &0.042$\pm$0.006 &0.047 & 0.034 & 0.33 
                               &0.052&0.030\\ \hline
   $\phi\pi^0$                 &$<6.8\times 10^{-4}$& 0 & 0 & 0 
                               &0&0\\ \hline
   $\rho^0\eta$                &0.0193$\pm$0.0023 &0.0174 & 0.0104 & 0.0186 &
                               0.0140&0.0144   \\ \hline
   $\omega\eta$                &0.158$\pm$0.016 &0.131 & 0.129 & 0.146 
                               &0.146&0.150\\ \hline
   $\phi\eta$                  &0.065$\pm$0.007 &0.065 & 0.062 & 0.076
                               &0.064&0.058\\ \hline
   $\rho^0\eta'$                 &0.0105$\pm$0.0018& & & &
                               0.0098 & 0.0112 \\ \hline
   $\omega\eta'$               &0.0168$\pm$0.0025 & & & &
                               0.167 & 0.0169 \\ \hline
   $\phi\eta'$                 &0.033$\pm$0.004 & & & &0.033&0.032\\ \hline
   $\chi^2$                   & &4.8 &57 & 21 & 8.2 & 43

\end{tabular}
\label{table:2}
\end{table}
 
\end{document}